\documentclass[sn-basic]{sn-jnl}

\jyear{2021}%

\theoremstyle{thmstyleone}%
\theoremstyle{thmstyletwo}%

\theoremstyle{thmstylethree}%

\begin{document}

\title[A Review on the Optimal Fingerprinting Approach in Climate Change Studies]{A Review on the Optimal Fingerprinting Approach in Climate Change Studies}


\author[1]{\fnm{Hanyue} \sur{Chen}}

\author*[1]{\fnm{Song Xi} \sur{Chen}}\email{songxichen@pku.edu.cn}

\author[2]{\fnm{Mu} \sur{Mu}}\email{mumu@fudan.edu.cn}

\affil[1]{\orgdiv{Center for Statistical Science}, \orgname{Peking University}, \orgaddress{
\city{Beijing}, 
\country{China}}}

\affil[2]{\orgdiv{Department of Atmospheric and Oceanic Sciences}, \orgname{Fudan Univeristy}, \orgaddress{
\city{Shanghai}, 
\country{China}}}


\abstract{We provide a review on the "optimal fingerprinting" approach as summarized in Allen and Tett (1999) from a point view of statistical inference in light of the recent criticism of McKitrick (2021).   
Our review finds that the "optimal fingerprinting" approach would survive much of McKitrick (2021)'s criticism under two conditions: (i) the null simulation of the climate model is independent of the physical observations and (ii) the null simulation provides consistent estimation of the residual covariance matrix of the physical observations,  both depend on the conduction and the quality of the climate models.  
If the latter condition fails, the estimator would be still unbiased and consistent under routine conditions, but losing the "optimal" aspect of the  approach. 

The residual consistency test suggested by Allen and Tett (1999) is valid for checking the agreement between the residual covariances of the null simulation and the physical observations. 
We further outline the connection between the “optimal fingerprinting" approach and the Feasible Generalized Least Square. 

}




\maketitle

\section{Introduction}\label{sec1}
The optimal fingering printing approach proposed in a series of papers starting from 
\cite{AT99} (herein AT99) followed by \cite{AS03} and \cite{SA03/2}), has been the corner stone in the detection and assessment of climate change due to human activities in the last 20 years, and has been the adopted methods by numerous Intergovernmental Panel on Climate Change (IPCC) 
 reports on the climate change. 
Recently, \cite{M21} (M21) offered a very critical assessment on  
\cite{AT99},  suggesting there were serious problems with the approach. 

The criticism of \cite{M21} is well summarized in its abstract "AT99 stated the GM Theorem incorrectly, omitting a critical condition altogether, their GLS method cannot satisfy the GM conditions, and their variance estimator is inconsistent by construction. Additionally, they did not formally state the null hypothesis of the RCT (residual consistency test) nor identify which of the GM conditions it tests, nor did they prove its distribution and critical values, rendering it uninformative as a specification test.”

%
%

This paper provides a statistical review on AT99  in light of the critics of \cite{M21}, and makes remarks on the key aspects of the AT99 formulation.  

We provide conditions under which the optimal fingerprint approach would stand, which are largely related to the null simulation of the climate models and the accuracy of the covariance  offered by the climate models as approximation to the residual covariance of the observations. 


\section{Review on AT99's Formulation}\label{sec2}

The main idea of AT99 is to utilize the Gauss-Markov (GM) theorem for the homogeneous linear regression model to make the detected fingerprints “optimal”, which means in the statistical terms,  to obtain estimates of the regression coefficients with the smallest variation so as to  attain the largest signal (regression coefficient estimates) to noise (standard deviation of the estimates) ratio. AT99 tried to  
utilize the GM theorem's ability to generate the Best linear unbiased estimator (BLUE) for the regression coefficients. The BLUE property would ensure the best signal to noise ratio in the detected human fingerprints on the climate change.  

The road to BLUE is not a direct one, as the homogeneous variation conditions for the GM theorem  were not readily  satisfied. 
AT99 suggested an approach that is well connected to the feasible generalized regression as pointed out by \cite{M21} with an approach that is uncommon to statisticians and econometricians in  estimating the residual covariance matrix  base on separate simulation of climate models.  

\subsection{Model and Key Aspect of Estimation} \label{sec2.1}

The Linear regression model considered in AT99 is 
\begin{equation}
\mathbf{y}=\mathbf{X} \boldsymbol{\beta}+\mathbf{u}\label{AT1}\tag{AT1}
\end{equation}
where $\mathbf{y}$ is the $\ell$-dimensional  vector of observations on certain climate variables, $\mathbf{X}$ is the design matrix of $\ell \times m$ corresponding to $m$ (climate) model-simulated response-patterns, which are commonly called covariates in statistics, and are obtained from ensembles of climate models.  
Throughout this review, the original equation numbers of AT99 are labeled as (AT.numeric number), while other equations are labelled separately. 

The $\mathbf{u}$ represents  the “climate noise” which was assumed to be multivariate normally distributed with the covariance matrix 
\begin{equation}
\mathbf{C}_{N} \equiv \mathcal{E} (\mathbf{uu}^{T}). \label{AT2}\tag{AT2}
\end{equation}
Here $\mathcal{E}$ denotes  the mathematical expectation on random variables in the physical world of observations.

\bigskip 
\noindent{\bf Remark 1.} AT99 did not present the regression identification condition
\begin{equation}
E(\mathbf{u}\mid\mathbf{X}) =0 \label{eq:iden1}
\end{equation}
which has drawn heavy criticism from \cite{M21}.  Our reading is that AT99 implicitly assumed such a condition.  
However, as will be shown in {\bf Remark 8} below, (\ref{eq:iden1}) is not enough for the suggested approach.

The major idea of AT99 is  to utilize the GM theorem for attaining the best  linear unbiased estimator(BLUE) of $\boldsymbol{\beta}$, which can attain the smallest mean square error  in the estimated $\boldsymbol{\beta}$-coefficients and hence the highest signal to noise ratio (SNR)  in the fingerprint detection. 

As $\mathbf{\mathbf{C}}_N \ne \sigma^2 \mathbf{I}_{\ell}$ due to the spatial and temporal heterogeneity in the residuals of the observations, to satisfy the variance condition that 
$$E(\mathbf{uu}^{T})= \sigma^2 \mathbf{I}_{\ell}$$
of the GM theorem,  AT99 suggested a pre-whitening operator via a $\kappa' \times \ell $ matrix $\mathbf{P}$ 
 such that
\begin{equation} 
\mathcal{E} \left(\mathbf{P u u}^{T} \mathbf{P}^{T}\right)=\mathbf{P C_{N}} \mathbf{P}^{T}=\mathbf{I}\label{AT3}\tag{AT3}
\end{equation} 

\bigskip
\noindent{\bf Remark 2.} AT99 was vague about the dimension of $\mathbf{P}$ and $\mathbf{I}$ in \eqref{AT3}, which should be $\mathbf{I}_{\kappa'}$ where $\kappa'$ is the rank of $\mathbf{P}$ and   $ \kappa' \le \ell$.  In the original GM theorem, $\kappa'=\ell$. 
There should be another restriction that $\kappa' > m$, the dimension of $\boldsymbol{\beta}$, 
in order to make  $\mathbf{X}^T \mathbf{P}^ T \mathbf{P} \mathbf{X}$ used in the generalized least square (GLS) full rank, which was only implicitly assumed in AT99 to ensure positive degree of freedom in the $\chi^2$-distribution in  (AT18). 

Ignoring the randomness of $\mathbf{P}$ for the moment, if $\mathbf{P}$ is known, 
Gauss-Markov theorem would imply that by left multiplying  $\mathbf{P}$ on all both sides of (\ref{ATeq:cond1}),  the ordinary least square (OLS) estimator $\boldsymbol{\beta}$  on the rotated data is 
\begin{equation}
\begin{aligned}
\tilde{\boldsymbol{\beta}}&=\left(\mathbf{X}^{T} \mathbf{P}^{T} \mathbf{P} \mathbf{X}\right)^{-1} \mathbf{X}^{T} \mathbf{P}^{T} \mathbf{P y}\\
&=\left(\mathbf{X}^{T} \mathbf{C}_{N}^{-1} \mathbf{X}\right)^{-1} \mathbf{X}^{T} \mathbf{C}_{N}^{-1} \mathbf{y} \equiv \mathbf{F}^{T} \mathbf{y}
\end{aligned}
\label{AT4}\tag{AT4} 
\end{equation}
where $\mathbf{F} = \left(\mathbf{X}^{T} \mathbf{P}^{T} \mathbf{P} \mathbf{X}\right)^{-1} \mathbf{X}^{T} \mathbf{P}^{T} \mathbf{P}$, and 
the covariance of $\tilde{\boldsymbol{\beta}}$  is 
\begin{equation}
Var(\tilde{\boldsymbol{\beta}})=\left(\mathbf{X}^{T} \mathbf{C}_{N}^{-1} \mathbf{X}\right)^{-1}.
\label{AT6}\tag{AT6}
\end{equation}  

\bigskip 
\noindent {\bf Remark 3.} The $\mathbf{\mathbf{C}}_N^{-1} $ is understood as the Moore–Penrose generalized inverse of $\mathbf{\mathbf{C}}_N$, which is $ \mathbf{P}^T \mathbf{P}$, in the case of $\kappa' < \ell$.  A more general expression for the variance is 
\begin{equation}
Var(\tilde{\boldsymbol{\beta}})=\left(\mathbf{X}^{T}  \mathbf{P}^{T} \mathbf{P} \mathbf{X}\right)^{-1} 
\label{eq:AT6a} 
\end{equation} 
 
{A key formulation of AT99 is on estimation of $\mathbf{\mathbf{C}}_N$ via separate simulations of certain climate model that would be consistent to the null setting of no human forcing on the climate, namely the so-called null setting in the climate models. 
Specifically,  }  
$\mathbf{C}_{N}$ is estimated by 
\begin{equation} 
\hat{\mathbf{C}}_{N}=\frac{1}{n} \mathbf{Y}_{N} \mathbf{Y}_{N}^{T}\label{AT12}\tag{AT12}
\end{equation}
where the columns of $\mathbf{Y}_{N}$ ($\ell\times n$ matrix)  represent $n$ response vectors of size $\ell$ from $n$ ensembles of the simulation runs of the null setting.

\bigskip
\noindent{\bf Remark 4.}  This external way of estimating the covariance matrix $\mathbf{\mathbf{C}}_N$  was quite different from the internal way commonly practised in statistics and econometrics.  {\bf The $n$ ensemble runs should be mutually independent of each other, and should be independent to the design matrix $\mathbf{X}$ in \eqref{AT1}. } 
A merit of this external estimation of $\mathbf{\mathbf{C}}_N$ is to make the estimation of the error covariance separate from the potential regression model misspecification (for instance omitted variables and  non-linearity).  
Let $\tilde{\mathbf{C}}_N$ be the the underlying covariance matrix of the ensembled null simulations to which $\hat{\mathbf{C}}_N$ converges to in probability.  
A drawback of this external approach is that it may be  challenging to make 
$\tilde{\mathbf{C}}_N$ match $\mathbf{\mathbf{C}}_N$ of the physical world. This is actually the most important factor that determines the estimated fingerprint $\tilde{\boldsymbol{\beta}}$ being BLUE or not. 

\bigskip 
\noindent{\bf Remark 5. }  A key requirement needed here is that the null simulation experiments should be statistically independent of the observed physical world 
governing Model \eqref{AT1} in order to ensure the key identification condition (\ref{ATeq:cond1}) 
 in {\bf Remark 8}. As will be shown later, this independence would be the key to ensure $\tilde{\boldsymbol{\beta}}$ being unbiased. 

Due to the limitation of the ensemble runs of the climate models,
$\hat{\mathbf{C}}_{N}$'s rank is only a $\kappa < \ell$, making $\hat{\mathbf{C}}_{N}$ non-invertible.  By  resorting to the generalized inverse  of $\hat{\mathbf{C}}_{N}$, $\mathbf{P}$ can be  approximated by a $\mathbf{P}^{(\kappa)}$. 
{Specifically, support  $\hat{\mathbf{C}}_N$ 
 admits the spectral decomposition that $\hat{\mathbf{C}}_N=\sum_{i=1}^r \lambda_i \nu_i \nu_i^T$ where ${\lambda}_{i=1}^{\kappa}$ are the positive eigen-values of $\hat{\mathbf{C}}_N$ ranked in descending order 
and ${\nu}_{i=1}^{\kappa}$ are the corresponding eigen-vectors.  One way to attain the $\kappa \times \ell$ matrix $\mathbf{P}^{(\kappa)}$  
 is to define } 
\[
\mathbf{P}^{(\kappa)} = (\lambda_1^{-1/2} \nu_1, \dots, \lambda_k^{-1/2} \nu_k)^T 
\]
Then, it can be verified that 
\begin{equation}
\mathbf{P}^{(\kappa)} \hat{\mathbf{C}}_N {\mathbf{P}^{(\kappa)}}^T = \mathbf{I}_{\kappa}
\end{equation}

\bigskip 
\noindent{\bf Remark 6.} {The reduced rank of 
$\hat{\mathbf{C}}_{N}$ implies that $\hat{\mathbf{C}}_{N}$ may not be a consistent estimator of  $\mathbf{C}_{N}$. 
This is an aspect that receives the bulk of the criticism in \cite{M21}. 
AT99 seems to suggest that  $\hat{\mathbf{C}}_{N}$ was designed to consistently estimate $\sum_{i=1}^k \lambda_i \nu_i \nu_i^T$, the first $\kappa$-terms in the spectral decomposition of $\mathbf{\mathbf{C}}_N$.  How to achieve this statistically is unclear to us.  } 

\bigskip 

\noindent{\bf Remark 7.}  Our understanding is that  $\hat{\mathbf{C}}_{N}$ is a consistent estimator of $\tilde{\mathbf{C}}_N$,  
the underlying covariance matrix of the null simulation of the climate model, which may be different from $\mathbf{\mathbf{C}}_N$ due to the limitation of the climate model.   If $\tilde{\mathbf{C}}_N \ne \mathbf{\mathbf{C}}_N$ as suspected by \cite{M21}, the estimator $\tilde{\boldsymbol{\beta}}$ is still consistent although will not have the smallest variance, hence giving up the BLUE property but still unbiased and consistent under the independence condition as we will show shortly. 


\bigskip 

\noindent {\bf Remark 8.}  In order to employ the GM theorem, 
the null climate experiments used to attain $\hat{\mathbf{C}}_N$ should be independent of  
the observed physical "experiments" that governs  Model \eqref{AT1}. If the independence can be ensured, 
which should be satisfied as the simulation runs should be statistically independent of the physical world.  Then,  $\mathbf{P}^{(\kappa)}$ matrix 
would be independent of $\mathbf{u}$ and $\mathbf{X}$ in  \eqref{AT1}, and then 
\begin{equation} 
 E(\mathbf{u} \mid\mathbf{X}, \mathbf{P}^{(\kappa)}) = E(\mathbf{u}\mid\mathbf{X}) =0. \label{ATeq:cond1} 
\end{equation} 
The statistical meaning of (\ref{ATeq:cond1}) is actually the conditional independence between $\mathbf{u}$ and $\mathbf{P}^{(\kappa)}$ given $\mathbf{X}$, slightly weaker than the full independence as mentioned above. 
{AT99 did not explicitly present this most primitive regression identification condition 
$E(\mathbf{u}\mid\mathbf{X}) =0$, neither mention (\ref{ATeq:cond1}) above, which had caused severe criticism from \cite{M21}.}

It is readily shown that under (\ref{ATeq:cond1}), as $E(\mathbf{y} \mid \mathbf{X}, \mathbf{P}^{(\kappa)}) = \mathbf{X}\boldsymbol{\beta}$, 
\begin{eqnarray}
&& E(\tilde{\boldsymbol{\beta}}) = E(E( \tilde{\boldsymbol{\beta}} \mid \mathbf{X}, \mathbf{P}^{(\kappa)})) \nonumber \\ 
&=&  E\left(\mathbf{X}^{T} \mathbf{P}^{(\kappa)^T} \mathbf{P}^{(\kappa)} \mathbf{X}\right)^{-1} \mathbf{X}^{T} \mathbf{P}^{(\kappa)^T} \mathbf{P} E(\mathbf{y} \mid\mathbf{X}, \mathbf{P}^{(\kappa)}) \nonumber \\ 
&=& \boldsymbol{\beta} \label{eq:unbiased}
\end{eqnarray}
indicating $\tilde{\boldsymbol{\beta}}$ being unbiased and having addressed the unbiased issue concerned by \cite{M21}.

\subsection{BLUE Property} \label{sec2.2}

We now consider the BLUE property offered by rotating the data by $\mathbf{P}^{(\kappa)}$, a key aspect of AT99 approach.  We assume  (\ref{ATeq:cond1}) and $\tilde{\mathbf{C}}_N=\mathbf{C}_N$, and show that  $\tilde{\boldsymbol{\beta}}$ is a restricted BLUE and is a full BLUE if $\kappa = \ell$. 


Let  us  consider another arbitrary linear unbiased estimator $\hat{\boldsymbol{\beta}}$ of $\boldsymbol{\beta}$ such that  $\hat{\boldsymbol{\beta}}=\mathbf{A} \mathbf{P}^{(\kappa)}\mathbf{y}$ for an arbitrary $m \times \kappa$ constant matrix $\mathbf{A}$, 
and 
$$\mathbf{D} = \mathbf{A} - (\mathbf{X}^T \mathbf{P}^{(\kappa) T} \mathbf{P}^{(\kappa)} \mathbf{X})^{-1}\mathbf{X}^T \mathbf{P}^{(\kappa) T}.$$  
The unbiasedness of $\hat{\boldsymbol{\beta}}$ suggests that 
\begin{equation}
\begin{aligned}
E(\hat{\boldsymbol{\beta}}\mid\mathbf{X}, \mathbf{P}^{(\kappa)})&=E(\mathbf{A} \mathbf{P}^{(\kappa)} y\mid\mathbf{X}, \mathbf{P}^{(\kappa)})\\
&=\boldsymbol{\beta} +\mathbf{D} \mathbf{P}^{(\kappa)} \mathbf{X} \boldsymbol{\beta}= \boldsymbol{\beta} 
\end{aligned}
\end{equation} 
implying that $\mathbf{D} \mathbf{P}^{(\kappa)} \mathbf{X}\boldsymbol{\beta} =0$. 

It can be shown that  
\begin{eqnarray}
&& Var(\hat{\boldsymbol{\beta}}\mid\mathbf{X},\mathbf{P}^{(\kappa)}) \nonumber \\ 
&=&E(\mathbf{A}\mathbf{P}^{(\kappa)}\mathbf{uu}^T{\mathbf{P}^{(\kappa)}}^T\mathbf{A}^T\mid\mathbf{X},\mathbf{P}^{(\kappa)}) \nonumber \\
&=& \mathbf{D} \mathbf{D}^T+ Var(\tilde{\boldsymbol{\beta}}\mid\mathbf{X},\mathbf{P}^{(\kappa)}).  \label{eq:varBLUE} 
\end{eqnarray}
As $\mathbf{D} \mathbf{D}^T\geq0$, $Var(\hat{\boldsymbol{\beta}}\mid\mathbf{X},\mathbf{P}^{(\kappa)})\geq Var(\tilde{\boldsymbol{\beta}}\mid\mathbf{X},\mathbf{P}^{(\kappa)})$. 

\bigskip 

\noindent{\bf Remark 9a.}  Note here the class of the linear estimators is restricted by $\mathbf{P}^{(\kappa)}$ when $\kappa < \ell$.   The demonstration above shows that the AT99 formulation with the enforcement (\ref{ATeq:cond1}) would make $\tilde{\boldsymbol{\beta}}$ is {\bf a restricted BLUE}, which would be a full unrestricted BLUE if $\kappa =\ell$.  

\noindent{\bf Remark 9b.} The largest threat to the BLUE property of $\tilde{\boldsymbol{\beta}}$ is $\tilde{\mathbf{C}}_N \ne \mathbf{\mathbf{C}}_N$, which would render the minimum variance aspect of $\tilde{\boldsymbol{\beta}}$ although it is still unbiased and consistent under the routine conditions from the perspective of feasible GLS (\cite{W02}).

\subsection{Model Inconsistency Check} \label{sec2.3}

AT99 was aware of the importance of $\tilde{\mathbf{C}}_N$ being a good approximation of $\mathbf{C}_N$ or not. 
Section 4 of AT99 is devoted to checking for this by examining  the residual inconsistency via a $\chi^2$ or a F-test. 
The null hypothesis $\cal{H}_0$ was worded as follows: 

"Our null-hypothesis, $\mathcal{H}_{0}$, is that the control simulation of climate variability is an adequate representation of variability in the real world in the truncated statespace which we are using for the analysis, i.e. the subspace defined by the first $\kappa$ EOFs of the control run does not include patterns which contain unrealistically low (or high) variance in the control simulation of climate variability. Because the effects of errors in observations are not represented in the climate model, $\mathcal{H}_{0}$ also encompasses the statement that observational error is negligible in the truncated state-space (on the spatio-temporal scales) used for detection. A test of $\mathcal{H}_{0}$, therefore, is also a test of the validity of this assumption."  

\cite{M21} complained that "they did not formally state the null hypothesis of the RCT nor identify which of the GM conditions it tests, nor did they prove its distribution and critical values, rendering it uninformative as a specification test". 

\bigskip 

\noindent {\bf Remark 10.} The null hypothesis $\mathcal{H}_{0}$ in AT99 is actually $\mathcal{H}_0: \tilde{\mathbf{C}}_N = \mathbf{\mathbf{C}}_N= Var(\mathbf{u})$ where $\tilde{\mathbf{C}}_N$ is the underlying covariance matrix of the null simulation via the climate model representing "the control simulation of climate variability", and $\mathbf{\mathbf{C}}_N= Var(\mathbf{u})$ is "the  variability in the real world".   As having stated in {\bf Remark 7}, 
 $\hat{\tilde{\mathbf{C}}}_N$ would be a consistent estimator of $\tilde{\mathbf{C}}_N$ such that $\hat{\tilde{\mathbf{C}}}_N^{-1} = \tilde{\mathbf{C}}_N^{-1} + o_p(1)$ where $o_p(1)$ represents a term that converges to 0 in probability, and $\tilde{\mathbf{C}}_N^{-1}=\tilde{\mathbf{P}}^T \tilde{\mathbf{P}}$ is the generalized inverse. The latter is implicitly assumed by AT99 as shown in the following derivation for the $\chi^2_{\kappa-m}$ distribution. 

Under $\mathcal{H}_{0}:\tilde{\mathbf{C}}_N = \mathbf{\mathbf{C}}_N$,  the test statistic 
\begin{eqnarray} 
&& r^2 = \tilde{\mathbf{u}}^T \hat{\mathbf{C}}_N^{-1} \tilde{\mathbf{u}} \nonumber \\ 
&= & (\mathbf{Y}-\mathbf{X}\tilde{\boldsymbol{\beta}})^T\hat{\tilde{C}}_N^{-1}(\mathbf{Y}-\mathbf{X}\tilde{\boldsymbol{\beta}})\nonumber \\
&= & \mathbf{Y}^T(\mathbf{I}-\mathbf{F} \mathbf{X}^T) \tilde{\mathbf{C}}_N^{-1}(\mathbf{I}-\mathbf{X}\mathbf{F}^T)\mathbf{Y} + o_p(1) \nonumber \\
&= & {\mathbf{u}^*}^T A {\mathbf{u}^*}+o_p(1) \nonumber 
\end{eqnarray} 
where $\mathbf{u}^*=\tilde{\mathbf{P}}^{(\kappa)} \mathbf{u} \sim N(0,\mathbf{I}_\kappa)$  and  under $\mathcal{H}_0$ 
$$\mathbf{A}= \mathbf{\mathbf{C}}_N^{\frac{1}{2}}(\mathbf{I}_\kappa-\mathbf{F}\mathbf{X}^T)\mathbf{\mathbf{C}}_N^{-1}(\mathbf{I}_{\kappa}-\mathbf{X}\mathbf{F}^T)\mathbf{\mathbf{C}}_N^{\frac{1}{2}}.$$ 

It can be shown that $\mathbf{A}^2=\mathbf{A}$ and $rank(\mathbf{A})= trace(\mathbf{A}) =\kappa - m$.  Thus, $r^{2} \sim \chi_{\kappa-m}^{2}$ as indicated in (AT18).  

The key that  leads to $A$ being idempotent (the key for the $\chi^2$ distribution) is the assertion of $\tilde{\mathbf{C}}_N = \mathbf{\mathbf{C}}_N$ under $\mathcal{H}_{0}$. 

\bigskip 
\noindent {\bf Remark 11.} AT99 discussed on the power of the $\chi^2$-test via the spectral decomposition (EOF) of $\hat{\mathbf{C}}_N$. A formal power analysis can be conducted as follows. Under the alternative hypothesis $\mathcal{H}_1: \tilde{\mathbf{C}}_N \ne \mathbf{\mathbf{C}}_N$, 
\begin{eqnarray} 
r^2 &= & {\mathbf{u}^*}^T \tilde{\mathbf{A}} {\mathbf{u}^*}+o_p(1) \label{eq:RCTH1} 
\end{eqnarray} 
where $\tilde{\mathbf{A}} = \mathbf{\mathbf{C}}_N^{\frac{1}{2}}(\mathbf{I}-\tilde{\mathbf{F}}\mathbf{X}^T)\tilde{\mathbf{C}}_N^{-1}(\mathbf{I}-\mathbf{X}\tilde{\mathbf{F}}^T)\mathbf{\mathbf{C}}_N^{\frac{1}{2}}$
and $\tilde{\mathbf{F}}$ is the same as $F$ defined earlier except that it uses $\tilde{\mathbf{P}}$ to replace $\mathbf{P}$. 
As $\tilde{\mathbf{A}}$ is no longer idempotent under $\mathcal{H}_1$, $r^2$ is no longer $\chi^2$ distributed. Let $\{\lambda_j\}_{j=1}^{\kappa}$ be the eigenvalues of $\tilde{\mathbf{C}}_N$ and $\{\chi_{1 j}^2\}_{j=1}^{\kappa}$ be the independent $\chi_1^2$ random variables, then it can be shown that 
$r^2 \sim \sum_{j=1}^{\kappa} \lambda_j \chi_{1 j}^2$, 
which is a weighted independent $\chi_1^2$ distribution and can be used to evaluate the power of the $\chi^2$-test.  If $\hat{\mathbf{C}}_N$ can be consistently estimated,  the eigen-values of $\tilde{\mathbf{A}}$ can be numerically computed and then the power of the RCT can be calculated under the (\ref{eq:RCTH1}). 

\bigskip 
\noindent{\bf Remark 12.} It is unclear on the powerfulness of the $\chi^2$ test in detecting differences in the underlying variance of the null simulation and the variance of the residuals in the observed physical world. 
In particular, its relative performance to the likelihood ratio test 
(\cite{AndersonBook}, p.412) should be investigated as $\mathcal{H}_0$ is vital 
for the BLUE and the optimal fingerprinting detection approach.   


\section{Connection to Feasible Generalized Least Square} 

\cite{M21} pointed out that the approach of AT99 is connected to the  Feasible Generalized Least Square (FGLS) in econometrics and statistics (\cite{Carroll82}; \cite{W02}).  We would  spell out the link first and then make some comments. 

Like AT99, the FGLS is a two-stage procedure with the first stage obtaining 
an initial and yet consistent estimator $\hat{\boldsymbol{\beta}}_1$ of $\boldsymbol{\beta}$ (for instance the OLS without data rotation) based on Model \eqref{AT1}. Let 
\begin{equation}
\hat{u}_i = y_i - x_i \hat{\boldsymbol{\beta}}_1 \quad \mbox{for $i=1, \cdots, \ell$}.  \label{eq:estu} 
\end{equation}
Then, if Model \eqref{AT1} is valid, a consistent estimator of $\mathbf{\mathbf{C}}_N$ is  
\begin{equation}
\hat{\mathbf{C}}_{N 1} = \ell^{-1} \sum_{i=1}^{\ell} \hat{u}_i \hat{u}_i^T. \label{eq：hatCN1} 
\end{equation}
The second stage of FGLS is to rotate the data with the square root matrix $\mathbf{P}_1$ of $\hat{\mathbf{C}}_{N 1}^{-1}$,  as did for $\hat{\mathbf{C}}_N^{-1}$ with $\mathbf{P}^{(\kappa)}$, and attain the FGLS estimator 
\begin{equation}
\tilde{\boldsymbol{\beta}}_{FGLS} = (\mathbf{X}^T \hat{\mathbf{C}}_{N 1}^{-1} X)^{-1} \mathbf{X}^T \hat{\mathbf{C}}_{N 1}^{-1} \mathbf{Y}  \label{eq:fgls} 
\end{equation}
or in the notation of AT99 
\begin{equation}
\tilde{\boldsymbol{\beta}}_{FGLS} = (\mathbf{X}^T \mathbf{P}_{N 1}^T \mathbf{P}_{N 1} \mathbf{X})^{-1} \mathbf{X}^T  \mathbf{P}_{N 1}^T \mathbf{P}_{N 1} \mathbf{Y}.  \label{eq:fgls1} 
\end{equation}

\bigskip 
\noindent  {\bf Remark 13.}  Comparing the above FGLS with that of AT99, 
 the latter  may be viewed as a FGLS. The difference between the two is rested in the way of finding the estimates of  $\mathbf{\mathbf{C}}_N$. The FGLS does it internally with the observed data under Model \eqref{AT1}, while AT99 does it externally via separate simulation of climate models.  An drawback of the internal approach is that it would be more adversely impacted by the mis-specification of Model \eqref{AT1}, namely the estimated residual $\tilde{\mathbf{u}}$ would be impacted by the estimation in the regression part.  In contrast, the challenge of the AT99's external approach is rested in if $\tilde{\mathbf{C}}_N$ is indeed $\mathbf{\mathbf{C}}_N$, that is the very purpose of  the $\chi^2$ test. We believe the likelihood ratio test as mentioned in {\bf Remark 11} would be more powerful in detecting the differences and should be investigated.

\section{Conclusion}

Our review finds that the "optimal fingerprinting" approach is very much dependent on two key conditions, which would make the approach survive \cite{M21}'s criticism. One is the independent condition (\ref{eq:iden1}) and another is $\tilde{\mathbf{C}}_N =\mathbf{C}_N$. Both are related to how the null simulation of the climate model is conducted and how accurate the climate model is in approximating the real covariance $\mathbf{C}_N$ of the physical world. 
We would think the first condition would be easier to argue than the second one, and would believe that  as the enduring effort in building better and more accurate climate models continues, the difference between $\tilde{\mathbf{C}}_N$ and $\mathbf{C}_N$ would be decreased, and the fingerprint approach  would be  closer to be "optimal". 
%
The independence condition would provide the key justification to the  regression identification condition (\ref{ATeq:cond1}) in {\bf Remark 8}, which in turn ensures $\tilde{\boldsymbol{\beta}}$ is unbiased removing a key concern of \cite{M21}.  

The reduced rank of $\hat{\mathbf{C}}_N$ would not make $\tilde{\boldsymbol{\beta}}$ losing the BLUE property all together, but only make it a restricted BLUE as indicated in {\bf Remark 9},  provided $\tilde{\mathbf{C}}_N = \mathbf{\mathbf{C}}_N$. In the case of $\tilde{\mathbf{C}}_N \ne \mathbf{\mathbf{C}}_N$, $\tilde{\boldsymbol{\beta}}$  would still be a consistent and unbiased estimator under standard conditions of the linear regression if the independence between "the two worlds" is maintained by the simulations of the  climate models.  Our remarks made in \ref{sec2.1}-\ref{sec2.2} should address most of “The five assumptions” raised by \cite{M21}. 

The residual consistency test (RCT) proposed by AT99 is a valid test for checking on the agreement between the covariances ($\tilde{\mathbf{C}}_N$ vs $\mathbf{\mathbf{C}}_N$) of the two worlds as shown in \ref{sec2.3}, which means that a proper control of the probability of the type I error (rejecting when the underlying covariance matrix of the null simulation does match $\mathbf{\mathbf{C}}_N$ of  the physical world)  
can be realized.  As mentioned in {\bf Remarks 11} and {\bf 12}, its control on the probability of the type II error (not rejecting when the covariances of the two worlds are actually different) need further study.  

Moreover, we outline the connection between the "optimal fingerprinting" approach and the Feasible Generalized Least Square, which we hope would connect the field of climate change studies and the statistical approach for conducting generalized least square regression.   And more vigorous analysis on the asymptotic normality  of $\tilde{\boldsymbol{\beta}}$ can be carried out using the standard methods in asymptotic statistics with specific conditions, which would not to get into in this review. 

\backmatter

\bmhead{Acknowledgments}
We thanks Hongbin Lin and Shanshan Luo for assistance. 

\section*{Declarations}
\begin{itemize}
\item Funding
\\This research was supported by the National Natural Science Foundation of China (Grant No.92046021).
\item Conflicts of interest 
\\ There is no conflict of interest.  

\item Availability of data and material 
\\Not applicable. 

\item Code availability 
\\Not applicable.
\end{itemize}



\begin{thebibliography}{7}
\providecommand{\natexlab}[1]{#1}
\providecommand{\url}[1]{{#1}}
\providecommand{\urlprefix}{URL }
\providecommand{\doi}[1]{\url{https://doi.org/#1}}
\providecommand{\eprint}[2][]{\url{#2}}
 \bibcommenthead

\bibitem[{Allen and Tett(1999)}]{AT99}
Allen M, Tett S (1999) Checking for model consistency in optimal
  fingerprinting. Climate Dynamics 15:419--434. \doi{10.1007/s003820050291}

\bibitem[{Allen and Stott(2003)}]{AS03}
Allen MR, Stott PA (2003) Estimating signal amplitudes in optimal
  fingerprinting, part i: theory. Climate Dynamics 21(5-6):477--491.
  \doi{10.1007/s00382-003-0313-9}

\bibitem[{Anderson(2003)}]{AndersonBook}
Anderson TW (2003) An introduction to multivariate statistical analysis, 3rd
  edn. Wiley-Interscience, Hoboken, N.J

\bibitem[{Carroll(1982)}]{Carroll82}
Carroll RJ (1982) Adapting for heteroscedasticity in linear models. The Annals
  of Statistics 10(4):1224--1233. \doi{10.1214/aos/1176345987}

\bibitem[{McKitrick(2021)}]{M21}
McKitrick R (2021) Checking for model consistency in optimal fingerprinting: a
  comment. Climate Dynamics \doi{10.1007/s00382-021-05913-7}

\bibitem[{Stott et~al(2003)Stott, Allen, and Jones}]{SA03/2}
Stott P, Allen M, Jones G (2003) Estimating signal amplitudes in optimal
  fingerprinting. part ii: Application to general circulation models. Climate
  Dynamics 21:493--500. \doi{10.1007/s00382-003-0314-8}

\bibitem[{Wooldridge(2002)}]{W02}
Wooldridge J (2002) Econometric Analysis of Cross SEction and Panel Data, The
  MIT Press

\end{thebibliography}


\end{document}